\documentclass[final]{aipproc}
\layoutstyle{6x9}

\begin{document}

\title{Photon-jet coincidence measurements
in polarized $pp$ collisions at $\sqrt{s}=200$ GeV
with the STAR Endcap Calorimeter}




\author{Ilya Selyuzhenkov for the STAR Collaboration}
{ address= {Indiana University Cyclotron Facility,
2401 Milo B. Sampson Lane, Bloomington, IN 47408, USA}}

\begin{abstract}
Recent inclusive measurements with polarized proton-proton collisions at RHIC
provide significant constraints on the polarized gluon distribution, $\Delta g(x)$,
integrated over the gluon momentum range $0.02 < x < 0.3$.
Di-jet and photon-jet coincidence measurements will allow to study the $x$-dependence of $\Delta g(x)$.
In this report we present the status of photon-jet coincidence studies
for photons detected at forward pseudorapidity, $1.08<\eta<2$, using the STAR Endcap Calorimeter.
\end{abstract}

\maketitle

\paragraph{\bf Introduction}
Recent results from RHIC for the inclusive jet \cite{Abelev:2007vt} and
$\pi^0$ \cite{Adare:2007dg} double spin asymmetry $A_{LL}$ at mid-rapidity
provide significant constraints \cite{deFlorian:2009vb}
on the integral of the polarized gluon distribution
over the gluon momentum range $0.02 < x < 0.3$.
Information on the gluon spin distribution as a function of $x$, $\Delta g(x)$,
will add insight if the observed smallness
of the integral value originates from possible cancellations.
It is also of great importance to extrapolations over unmeasured regions of $x$.
Determining the $\Delta g(x)$ dependence
requires reconstruction of the initial-state parton kinematics,
 which can be achieved with di-jet or photon-jet coincidence measurements.
The photon-jet channel is dominated by a single partonic subprocess (quark-gluon Compton scattering),
and through measurement of the photon energy and direction,
along with the jet direction, allows more precise reconstruction
of the parton kinematics.
In this report we present the status of our photon-jet
coincidence studies for photons detected at forward rapidity,
$1.08<\eta<2$, using the STAR Endcap Electro-Magnetic Calorimeter (EMC) \cite{Allgower:2002zy}.
Full jet reconstruction at mid-rapidity, $|\eta|<0.8$,
is obtained using the STAR Time Projection Chamber (TPC)
\cite{Anderson:2003ur} and the Barrel EMC detectors \cite{Beddo:2002zx}.

\paragraph{\bf Data samples, event selection, and uncorrected yields}
Our analysis is based on
an integrated luminosity of 3.1pb$^{-1}$ 
longitudinally polarized proton-proton collisions at \mbox{$\sqrt{s}=200$ GeV}
which were recorded with the STAR detector during year 2006.
The trigger required at least one 2x2 cluster of towers in the Endcap EMC
with transverse energy greater than \mbox{5.2 GeV}
and transverse energy of the associated tower above \mbox{3.7 GeV}.
Two event samples (signal and background) of simulated $pp$ collisions
were used for detailed study of trigger and detector biases,
and to derive purity of the extracted signal.
Events were generated using PYTHIA 6.4 \cite{Sjostrand:2006za} with parameters
adjusted to CDF \mbox{"Tune A"} settings \cite{Field:2005sa}
in the parton transverse momentum range
2-25~GeV.
The signal rate is calculated based on simulated PYTHIA prompt photon production processes.
Physics backgrounds are studied with PYTHIA QCD hard scattering
two parton production processes.
Realistic response of the STAR detector is simulated with
the GEANT 3 Monte-Carlo package \cite{Geant3}
with full trigger emulation.
The generated luminosity for each Monte-Carlo sample is comparable to that of the data:
7pb$^{-1}$ for direct photon-jet, and 1pb$^{-1}$ for  QCD background samples.

All events were processed with the midpoint-cone jet finding algorithm \cite{Blazey:2000qt},
and only those with exactly two jets
which pointed back-to-back in azimuth, \mbox{$\cos(\phi_{jet}-\phi_{\gamma})< -0.8$},
were selected for the analysis.
The photon candidate is defined as the jet with the maximum neutral energy, and
its energy is calculated based on the 3x3 cluster of Endcap EMC towers
centered around the tower of highest energy.
Additional event fiducial cuts
on the photon candidate pseudo-rapidity $1.08< \eta_{\gamma}<2$
and transverse momentum \mbox{$p_{T}^{\gamma}>7$ GeV},
and on the away-side jet $|\eta_{jet}|<0.8$ and \mbox{$p_{T}^{jet}>5$ GeV} were applied.
\begin{figure}[h]
\includegraphics[height=0.28\textheight]{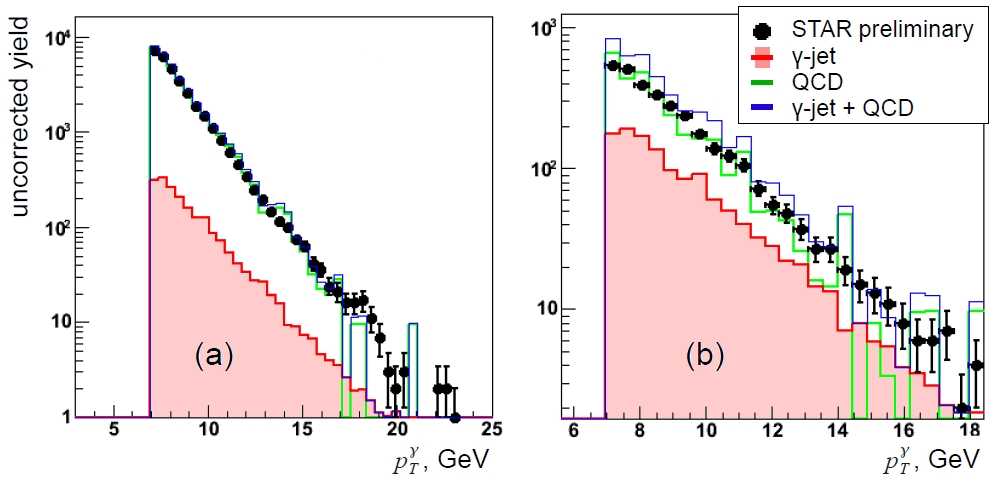}%
\label{Fig1}
\caption{
Uncorrected yields vs. photon candidate transverse momentum.
Black circles indicate the yield of photon-jet candidates from STAR run 6
$pp$ collisions at \mbox{$\sqrt{s}=200$ GeV},
red (green) lines show simulated prompt photons (QCD background) yield,
and blue lines represent sum of the simulated yields.}
\end{figure}
\mbox{Figure \ref{Fig1}(a)} shows acceptance and efficiency uncorrected yields for data and
simulation samples after applying event fiducial cuts.
The sum of the simulated yields were normalized to the yield in data.
Figure \ref{Fig1}(b) shows uncorrected yields after background suppression
according to the photon-jet isolation procedure discussed below.

\paragraph{\bf Transverse shower profile and photon-jet isolation}
The main source of physics background in the direct photon measurement
originates from multi-photon production processes, such as $\pi^0\to \gamma\gamma$ decay.
In this study, for multi-photon discrimination we used
the STAR Endcap Shower Maximum Detector (SMD) \cite{Allgower:2002zy}.
High SMD granularity allows for precise photon position reconstruction,
while SMD strip energy deposition
provides important information on the transverse electromagnetic shower profile.
For the shower shape study and cut optimization,
all events were pre-sorted into
four different categories based on
energy deposition in the two Endcap EMC pre-shower layers:
(a) $E_{pre1}=E_{pre2}=0$;
(b) $E_{pre1}=0$, $E_{pre2}>0$;
(c) $0 < E_{pre1}<4$ MeV; and
(d) $4 < E_{pre1}<10$ MeV.
Here $E_{pre1(2)}$ is the pre-shower energy deposition under the 3x3 cluster of towers.
\begin{figure}[h]
\includegraphics[height=0.28\textheight]{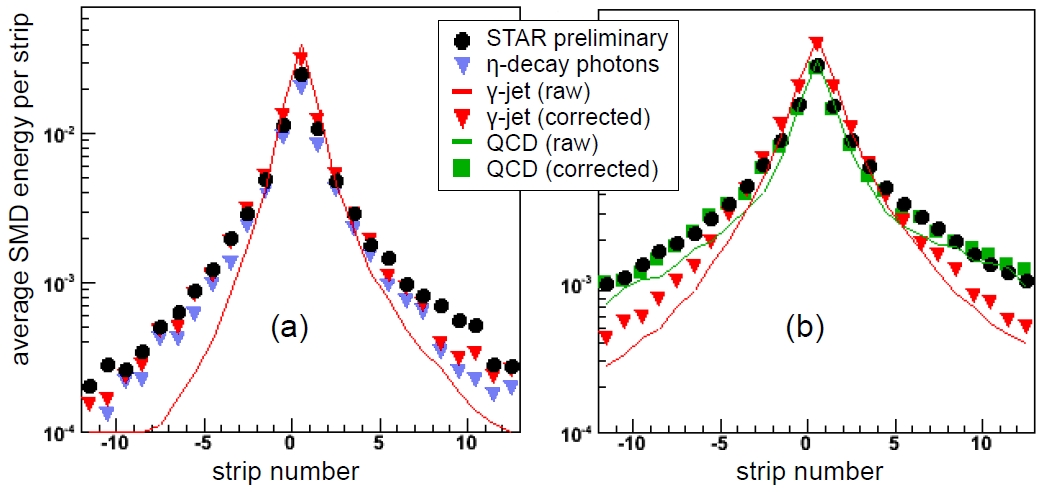}%
\label{Fig2}
\caption{
Photon candidate transverse shower profiles.
Black circles present shower shapes from real $pp$ collisions at $\sqrt{s} = 200$ GeV.
Red (green) lines correspond to simulated prompt photon (QCD background) shower shapes,
while red triangles (green squares) indicate shower profiles from data-driven simulations
(see text for details).}
\end{figure}
Figure \ref{Fig2} shows the average energy deposition per SMD strip vs.
relative distance to the strip of highest energy (strip number).
Figure \ref{Fig2}(a) shows the transverse energy profile for a
sample of events with $E_{pre1}=E_{pre2}=0$ (direct photon rich),
while \mbox{Fig. \ref{Fig2}(b)} presents a sample with $4 < E_{pre1}<10$ MeV (background dominated).
From \mbox{Fig. \ref{Fig2}} it is clear that
GEANT Monte-Carlo with the STAR run 6 geometry
generates transverse shower profiles which are
not consistent with those from the real data
(compare red or green line vs. black circles in Fig. \ref{Fig2}).
This disagreement has been fixed by event-by-event substitution
of the GEANT simulated SMD response for each of the Monte-Carlo photons
with that of isolated photons from real data.
To build a library of real photon shower shapes,
we used a sample of reconstructed $\eta$-mesons which decayed
into two well-separated photons such
that their showers do not overlap in SMD.
Shower shapes from this so-called data-driven simulation
are indicated in \mbox{Fig. \ref{Fig2}} by red triangles (direct photon)
and green squares (QCD background).
For comparison, the transverse shower profile of single photons
from $\eta$-meson decay is shown as blue triangles.

Direct photon shower shapes were parametrized
with a triple Gaussian function which was used to fit
the SMD response in data and simulation on an event-by-event basis.
For each event, the fit results are
subtracted from the observed SMD energy deposition, and the extra energy
on each side of the SMD peak is calculated.
The maximum of these energies (maximum sided-residual)
is further used for multi-photon background discrimination.
In addition, the following event information is used for background suppression:
(a) energy fraction of the 3x3 tower cluster
within a larger ($r=0.7$) radius in $\Delta\eta\times\Delta\phi$;
(b) number of Endcap EMC towers fired around photon candidate
within $r=0.7$;
(c) number of Barrel EMC towers fired around photon candidate within $r=0.7$;
(d) number of charged tracks reconstructed with the TPC around photon candidate within $r=0.7$;
(e) photon and jet transverse momentum balance, $[p_T^{\gamma}-p_T^{jet}]/p_T^{\gamma}$;
(f) post-shower energy deposition for the 3x3 cluster of Endcap EMC towers.
For this set of discriminating variables,
the cuts were optimized with a
Linear Discriminant Analysis (LDA) \cite{Hocker:2007ht}.
Given $i$ discriminating variables, $v_i$, the LDA
assigns a specific weight, $w_i$, and
all variables are combined into a
single linear discriminant: $D=\sum_i v_i w_i$.
The LDA optimizes weights in such a way that
signal and background discriminant distributions
are pushed as far as possible away from each other.
\begin{figure}[h]
\includegraphics[height=0.28\textheight]{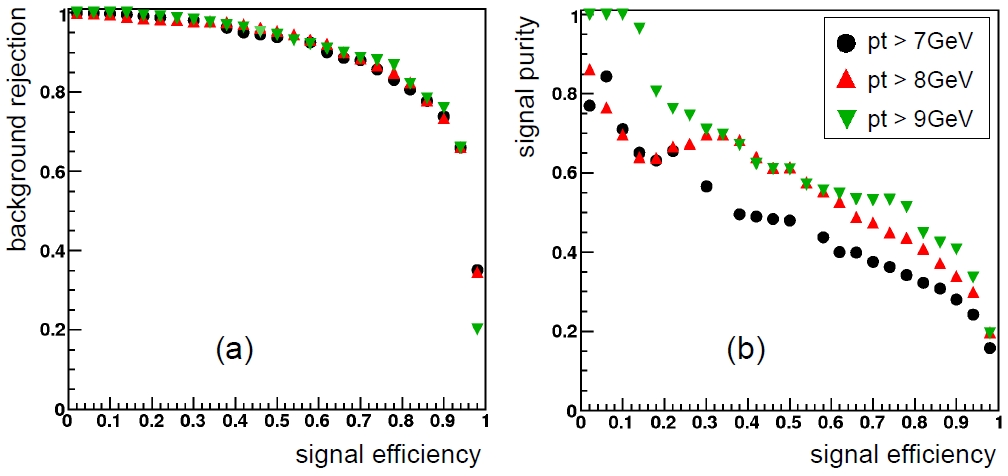}%
\label{Fig3}
\caption{
(a) Background rejection vs. signal efficiency. (b) Signal purity vs. signal efficiency.
Results are shown for the $E_{pre1}=0$, $E_{pre2}>0$ pre-shower condition.}
\end{figure}
Figure \ref{Fig3} shows the result of weight optimization with LDA for three different
cuts on the photon transverse momentum: $p_T^{\gamma}>7$, $8$, and $9$ GeV.
As can be seen from Fig. \ref{Fig3}(b), for a given efficiency of 70\% and the $E_{pre1}=0$, $E_{pre2}>0$ condition
we can reach 40\% purity in $p^{\gamma}_T>7$ GeV range.
As a function of $p^{\gamma}_T$,
signal purity averaged over all pre-shower conditions
varies between 25-40\% 
(corresponding uncorrected yields are show in Fig. \ref{Fig1}(b)).

\paragraph{\bf Summary}
The status of photon-jet coincidence studies
for photons detected at forward rapidity, $1.08<\eta<2$,  using the STAR detector is presented.
Depending on the photon transverse momentum,
we have reached 25-40\% overall purity of the photon-jet sample.
This result is an important step towards calculation of the total
photon-jet cross section in the photon rapidity range  $|\eta_{\gamma}|>1$.
With the statistics collected by STAR during this year's run 9,
this should allow for statistically significant double spin asymmetry measurements
with photon-jet coincidences.

\end{document}